\documentclass[%
 reprint,
 amsmath,amssymb,
 aps,
pra,
 longbibliography,
 lengthcheck,%
]{revtex4-1}

\usepackage{graphicx}
\usepackage{dcolumn}
\usepackage{bm}

\begin{document}

\preprint{APS/123-QED}

\title{Magnetism and Superconductivity in CeRhIn$_5$}

\author{Georg Knebel}
 \email{georg.knebel@cea.fr}
\author{Dai Aoki}%
 
\author{Jacques Flouquet}
 
\affiliation{
 Commissariat \`a l' \'Energie Atomique, INAC, SPSMS, 17 rue des Martyrs, 38054 Grenoble, France
}%


\date{\today}

\begin{abstract}
We give a review of recent results on the heavy fermion superconductor CeRhIn$_5$ which presents ideal conditions to study the interplay of antiferromagnetism and superconductivity. The main results are the appearance of domains of microscopic coexistence of antiferromagnetism and superconductivity under both, pressure and magnetic field scans. At zero magnetic field, under pressure coexistence appears from the onset of pressure induced superconductivity up to the critical pressure $p_c^\star$ where antiferrmognetism disappears abruptly. Under magnetic field, re-entrance of magnetic order inside the superconducting state occurs in the pressure range from $p_c^\star$ up to $p_c$. The antiferromagnetic quantum critical point is hidden by the onset of superconductivity. A striking point is the similarity of the phase diagram of CeRhIn$_5$ and that of the high $T_c$ cuprate superconductors where for the cuprates the carrier concentration replaces the pressure. Both systems are characterized by: hidden magnetic quantum criticality, field reentrance of magnetism and the strong link to a Fermi surface instability.

\end{abstract}

\pacs{Valid PACS appear here}
\maketitle


\section*{Introduction}

The discovery of new superconductors and the understanding of the mechanism that is responsible for the formation of the superconducting  state is central in modern solid state physics. Ideal model systems for this studies are so-called heavy fermion systems. These are $f$ electron systems in which the hybridization of the $f$electrons with the light conduction electrons give rise to the formation of heavy quasiparticles with an effective mass $m^\star$ up to 100 times larger than that of an usual metal. The superconducting pairing in these compounds is generally expected not to be a phonon mechanism but unconventional, i.e.~magnetic or valence fluctuations give rise to breaking of further symmetries. Up to now almost thirty different systems have been observed showing a superconducting transition either at ambient or under high pressure. 

The intense research on unconventional superconductivity (SC) in heavy-fermion systems started with the unexpected discovery of superconductivity in CeCu$_2$Si$_2$ in 1979 \cite{Steglich1979}. Later on in the 80$^{\rm th}$ superconductivity has been found in the U based heavy-fermion compounds. The development of high pressure experiments and the progress in sample quality led to the observation of pressure induced superconductivity in Ce based heavy fermion compounds close to their magnetic instability. The first examples are CeCu$_2$Ge$_2$ \cite{Jaccard1992}, CeRh$_2$Si$_2$ \cite{Graf1996}, CeIn$_3$ and CePd$_2$Si$_2$ \cite{Mathur1998}.
These compounds are antiferromagnetically ordered  and superconductivity appears only under high pressure and at very low temperature ($T_c<600$~mK), which makes a detailed analysis of the competition of both phenomena experimentally very difficult. 

\begin{figure}
\begin{center}
\includegraphics[width=0.6\hsize,clip]{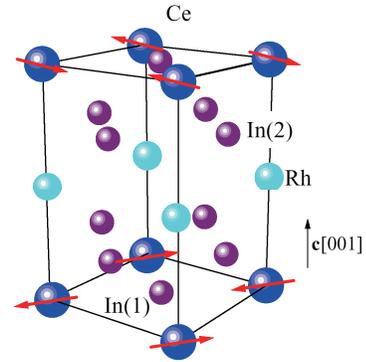}%
\caption{\label{crystal_structure} Crystal structure of CeRhIn$_5$. It contains planes of CeIn$_3$ which are separated by planes of RhIn$_2$. This gives rise to a quasi two-dimensional main Fermi surface with cylinders along the $c$ direction. The incommensurate antiferromagnetic structure is indicated by the arrows.}
\end{center} 
\end{figure}

A breakthrough was the discovery of pressure induced superconductivity  in antiferromagnetically ordered CeRhIn$_5$ in Los Alamos in 2000 and later at ambient pressure in CeCoIn$_5$ and CeIrIn$_5$.\cite{Hegger2000,Petrovic2001,Petrovic2001b}. In this so-called Ce-115 family of compounds due to their high  superconducting transition temperature of $T_c \approx 2$~K it is possible to perform precise measurements of the magnetic as well as of the superconducting properties and to study their interaction. The crystal structure of these compounds is tetragonal and is built of planes of CeIn$_3$ and planes of $M$In$_2$ ($M$ = Rh, Co, or Ir) stacked sequently along the $c$ axis (see Fig. \ref{crystal_structure}). This stacking reinforces the in-plane two dimensional interactions between the cerium ions. The enhancement of the two dimensional electronic character is considered to be the origin of the increase the superconducting transition temperature in the Ce-115 family \cite{Monthoux2001,Monthoux2002}.

Generally, the ground state of a Ce heavy-fermion system is determined by the competition of the indirect Ruderman Kittel Kasuya Yosida (RKKY) interaction which provokes magnetic order of localized moments mediated by the light conduction electrons and the Kondo interaction. This last local mechanism causes a paramagnetic ground state due the screening of the local moment of the Ce ion  by  the conduction electrons. Both interactions depend critically on the hybridization of the $4f$ electrons with the conduction electrons. High pressure is an ideal tool to tune the hybridization and the position of the 4$f$ level with respect to the Fermi level. Therefore high pressure experiments are ideal to study the critical region where both interactions are of the same order and compete. To understand the quantum phase transition from the antiferromagnetic (AF) state to the paramagnetic (PM) state is actually one of the fundamental questions in solid state physics. Different theoretical approaches exist to model the magnetic quantum phase transition such as spin-fluctuation theory of an itinerant magnet \cite{Moriya1985,Millis1993,Moriya1995}, or a new so-called 'local' quantum critical scenario \cite{Si2001,Gegenwart2008}.
Another efficient source to prevent long range antiferromagnetic order is given by the valence fluctuations between the trivalent and the tetravalent configuration of the cerium ions \cite{Miyake2007}.

The interesting point is that in these strongly correlated electron systems the same electrons (or renormalized quasiparticles) are responsible for both, magnetism and superconductivity. 
The above mentioned Ce-115 family is an ideal model system, as it allows to study both, the quantum critical behavior and the interplay of the magnetic order with a superconducting state. Especially, as we will be shown below, unexpected observations will be found, if a magnetic field is applied in the critical pressure region. 

\section*{Pressure-temperature phase diagram}
\begin{figure}[t]
\begin{center}
\includegraphics[width=0.6\hsize,clip]{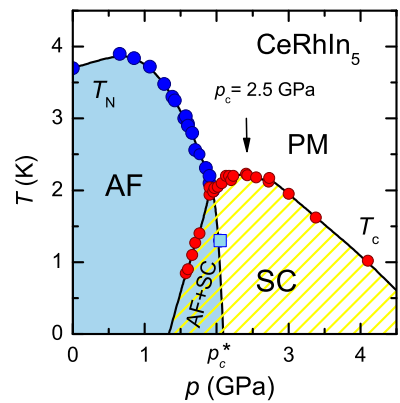}%
\caption{\label{CeRhIn5_PD} Pressure--temperature phase diagram of CeRhIn$_5$ at zero magnetic field determined from specific heat measurements with antiferromagnetic (AF, blue) and superconducting phases (SC, yellow). When $T_c < T_N$ a coexistence phase AF+SC exist. When $T_c > T_N$ the antiferromagnetic order is abruptly suppressed. The blue square indicate the transition from SC to AF+SC after Ref.~\onlinecite{Yashima2007}. }
\end{center} 
\end{figure}

In this article we concentrate on the compound CeRhIn$_5$.  At ambient pressure the RKKY interaction is dominant in CeRhIn$_5$ and magnetic order appears at $T_N = 3.8$~K. However, the ordered magnetic moment of $\mu = 0.59\mu_B$ at 1.9~K is reduced of about 30\% in comparison to that of Ce ion in a crystal field doublet without Kondo effect \cite{Raymond2007}. Compared to other heavy fermion compounds at $p = 0$ the enhancement of the Sommerfeld coefficient of the specific heat ($\gamma = 52$~mJ mol$^{-1}$K$^{-2}$) \cite{Knebel2004} and the cylotron masses of electrons on the extremal orbits of the Fermi surface is rather moderate \cite{Hall2001, Shishido2002}. The topologies of the Fermi surfaces of CeRhIn$_5$ are cylindrical and almost identical to that of LaRhIn$_5$ which is the non 4$f$ isostructural reference compound. From this it can be concluded that the 4$f$ electrons in CeRhIn$_5$ are localized and do not contribute to the Fermi volume \cite{Hall2001,Shishido2002}.

By application of pressure, the system can be tuned through a quantum phase transition. 
The N\'eel temperature shows a smooth maximum around 0.8~GPa and is monotonously suppressed for higher pressures.  However, CeRhIn$_5$ is also a superconductor in a large pressure region from about 1.3 to 5 GPa. It has been shown that when the superconducting transition temperature $T_c > T_N$ the antiferromagnetic order is rapidly suppressed (see figure \ref{CeRhIn5_PD}) and vanishes at a lower pressure than that expected from a linear extrapolation to $T=0$.  
Thus the pressure where $T_c = T_N$ defines a first critical pressure $p_c^\star$ and clearly just above $p_c^\star$ anitferromagnetism collapses. The intuitive picture is that the opening of a superconducting gap on large parts of the Fermi surface above $p_c^\star$ impedes the formation of long range magnetic order. A coexisting phase AF+SC in zero magnetic field seems only be formed if on cooling first the magnetic order is established. We will discuss below the microscopic evidence of an homogeneous AF+SC phase. 

At ambient pressure CeRhIn$_5$ orders in an incommensurate magnetic structure \cite{Bao2000} with an ordering vector of \boldmath $q_{ic}$\unboldmath =(0.5, 0.5, $\delta$) and $\delta=0.297$  that is a magnetic structure with a different periodicity than the one of the lattice. Generally, an incommensurate magnetic structure is not favorable for superconductivity with $d$ wave symmetry, which is realized in CeRhIn$_5$ above $p_c^\star$ \cite{Mito2001}. Neutron scattering experiments under high pressure do not give conclusive evidence of the structure under pressures up to 1.7~GPa which is the highest pressure studied up to now \cite{Llobet2004,Raymond2008,Aso2009}.  The result is that at 1.7~GPa the incommensurability has changed to $\delta \approx 0.4$. The main difficulty in these experiments with large sample volume is to ensure the pressure homogeneity. Near $p_c^\star$ the control of a perfect hydrostaticity is a key issue as the material reacts quite opposite on uniaxial strain applied along the $c$ and $a$ axis. 

From recent nuclear quadrupol resonance (NQR) data it is followed that the magnetic order gets commensurate ($\delta = 0.5$) above 1.7~GPa when bulk superconductivity appears under pressure \cite{Yashima2009}. Below $p_c^\star$ it is observed that the phase transition on cooling  from AF to AF+SC looks quite inhomogeneous but far below $T_c$, NQR shows that 
 the spin-lattice relaxation $1/T_1$ is homogeneous independent on the local site what is also a nice hint of the coexistence of both states below $p_c^\star$ \cite{Kawasaki2003,Yashima2007,Yashima2009}. This is consistent with the findings in systems where the antiferromagnetic order in CeRhIn$_5$ is suppressed by doping. In these cases SC seems at least to coexist with magnetic order if the vector is commensurate \cite{Christianson2005,Ohira2007}.

\begin{figure}
\begin{center}
\includegraphics[width=.6\hsize,clip]{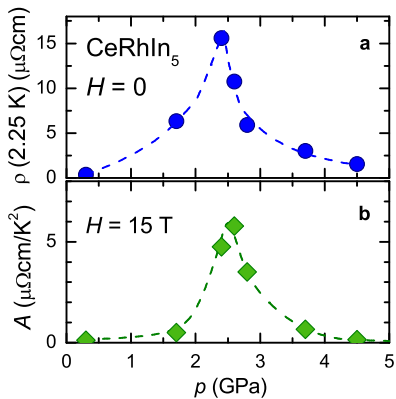}
\caption{\label{Acoeff_rho0} Pressure dependence of a) the resistivity at $T= 2.25$~K just above the superconducting transition, b) the $A$ coefficient of the resistivity measured at a field $H=15$T far above the upper critical field $H_{c2}$. Both quantities are strongly enhanced around the critical pressure $p_c$. (Lines are guides for the eye.)
}
\end{center} 
\end{figure}

Above $p_c^\star$ the ground state is a superconductor and magnetism is rapidly suppressed. Assuming a homogeneous AF+SC state the point in the $(p,T)$ phase diagram with $T_N = T_c$ is a tetra-critical point where four lines of second order phase transitions should meet and indeed, in the recent NQR experiment a magnetic transition inside the superconducting phase ($T_N<T_c$) has been observed just 0.1 GPa above $p_c^\star$(see Fig.~\ref{CeRhIn5_PD}) \cite{Yashima2007}. 
The actual proposal is that the onset of SC below $p_c^\star$ is associated with an incommensurate -- commensurate magnetic transition which may alter drastically the usual superconducting anomalies at $T_c$. It is interesting to note that NQR measurements also show that at $p=0$ far below the onset of the AF+SC state a tiny residual commensurate AF phase survives inside a dominant incommensurate ordered part leading to a possible parasitic SC signal even at ambient pressure \cite{Chen2006,Paglione2008}.


Where would be the magnetic quantum critical point of the system in absence of superconductivity? A linear extrapolation of $T_N$ to 0~K gives a critical pressure $p_c = 2.5$ GPa for the disappearance of magnetism which corresponds to the maximum of $T_c$. An indication of this critical pressure is a strong maximum in the resistivity at $T_c
=2.25$~K just above the onset of the superconducting transition due to the enhancement of the scattering caused by critical fluctuations \cite{Miyake2002} as shown in Fig.~\ref{Acoeff_rho0}a). The resistivity in this pressure range shows clearly a non-Fermi liquid behavior with a sub-linear temperature dependence \cite{Knebel2008, Park2008d}. However the deeper meaning of this unconventional temperature dependence is not completely understood. In the a spin-fluctuation scenario one would expect that the resistivity at the critical pressure would show a $T^{3/2}$ temperature dependence for a three dimensional antiferromagnetic system; maybe the dimensionality is reduced and but then a linear temperature dependence would be expected \cite{Moriya2003}. However, there a further theoretical proposals for a linear $T$ dependence of the resistivity such as critical valence fluctuations \cite{Holmes2004} or a selective Mott transition \cite{Pepin2008}. At least in our view the experimentally observed sub-linear $T$-dependence is much more an effect of an cross-over to the low temperature regime in difference to Ref.~\onlinecite{Park2008d}. Obviously $T_c$ appears at a higher temperature than the temperature $T_A$ where a $AT^2$ Fermi liquid law of the resistivity will be obeyed. 

\begin{figure*}[tb]
\begin{center}
\includegraphics[width=.8\hsize,clip]{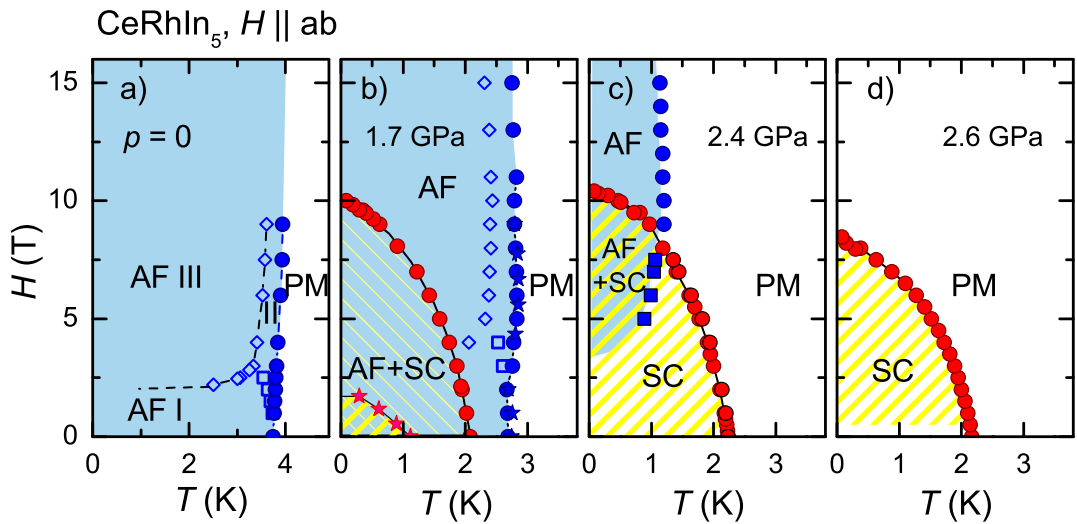}
\caption{\label{field} Magnetic field phase diagram of CeRhIn$_5$ at different pressures for a magnetic field applied in the $ab$ plane (blue symbols for magnetic transitions, red symbols for superconductivity). a-b) The magnetic phase diagram is almost unchanged compared to $p=0$ up to $p_c^\star$. However, $H_{c2} (T)$ detected by specific heat (red stars, taken from Ref.\onlinecite{Park2008b}) is much lower that detected by resistivity. This indicates the inhomogeneous superconducting state observed below $p_c^\star$. Remarkably magnetic order is induced inside the superconducting dome in the pressures $p_c^\star < p < p_c$, as shown for $p=2.4$~GPa in c). d) For $p> p_c$ no magnetic order appears.}
\end{center} 
\end{figure*}

Other indicators of the critical pressure are the strong maximum of the $A$ coefficient of the electronic resistivity $\rho = \rho_0 + AT^2$ measured at high magnetic fields of 15~T where superconductivity is suppressed.   
In difference to the prototypical heavy-fermion superconductor CeCu$_2$Si$_2$ where the magnetic quantum critical point ($p= 0$) and a sharp valence transition (at $p=4$~GPa) are clearly separated in pressure and two different superconducting domes exist \cite{Yuan2003, Holmes2004}, in most other heavy fermion superconductors such a clear separation of both criticalities is not possible. Thus, as function of pressure both criticalities seem to fall together, e.g.~in CeIn$_3$ \cite{Knebel2001}, CePd$_2$Si$_2$ \cite{Demuer2001}, and also here in CeRhIn$_5$ only one maximum in the pressure dependence of the $A$ coefficient occurs. The disappearance of magnetism at $p_c$ appears to be driven by the valence fluctuations.


\section*{Magnetic field effects}

The application of a magnetic field will lead to test the interplay between antiferromagnetism and superconductivity as $T_N (H)$ and $T_c (H)$ will respond differently to a field and thus the condition $T_c = T_N$ for the suppression of the long range order due to the onset of bulk superconductivity below $p_c$ will be modified. Furthermore, a new situation appears under magnetic field in a the superconducting state due to the creation of a mixed state with superconductivity and a vortex structure. 
 
 The magnetic properties at ambient pressure have been studied up to high fields of 60~T by magnetization measurements \cite{Takeuchi2001}. For a magnetic field $H \parallel c$ the magnetization increases monotonously and no anomaly can be detected. Thus the field to reach a polarized paramagnetic state at $H_m$ is above the experimental limits. In difference the magnetization measurements for $H \perp c$ two transitions have been observed at low temperatures at $H_{ic} \approx 2$~T and $H _m\approx 50$~T \cite{Takeuchi2001}. The low temperature magnetic phase diagram has been established by specific heat measurements showing three different magnetic phases \cite{Cornelius2001} and their magnetic structures have been determined by neutron scattering \cite{Raymond2007}. Figure~\ref{field} shows the $H-T$ phase diagrams for different pressures. At zero pressure the incommensurate structure (AF I) changes to commensurate (AF III) for a field of $H_{ic}\approx 2.25$~T with an ordering vector \boldmath $q$\unboldmath $=(0.5, 0.5, 0.25)$ and the magnetic structure has changed to a collinear one. The ordering in phase AF II is also incommensurate, but a collinear sine-wave structure is formed. 

As discussed previously the magnetic structure (at least in the condictions of the neutron scattering experiments) is incommensurate under pressures up to $p=1.7$~GPa even if the $z$ component of the magnetic propagation vector has changed to $\delta \approx 0.4$. Up to now no neutron scattering experiment exist for higher pressures and high magnetic fields thus the magnetic structures of different magnetic phases in Fig.~\ref{field} b) are not determined up to. The superconducting boundaris shown in Fig.~\ref{field} b) have been plotted from resistivity measurements what may not present the bulk superconducting phase and from ac calorimetry taken from Ref.\onlinecite{Park2008b}. (The pressure has been normalized to our measurements.) The temperature difference of the superconducting transitions derived from resistivity and specific heat in this pressure range is quite large \cite{Knebel2006}. The ac calorimetric studies performed in Los Alamos show that the upper critical field is significantly lower, e.g.~at $p=1.71$~GPa $H_{c2} (0) \approx 6$~T \cite{Park2008b}. Furthermore it should be noted that the upper critical field shows a very peculiar temperature dependence in the AF+SC state below $p_c^\star$ which requires future investigations \cite{Park2007b,Park2008b}. The upper critical field at $T=0$,  $H_{c2} (0)$ coincides with the field $H_{ic}$ indicating the change of the magnetic structure to a collinear one with $\delta = 0.25$, which increases with increasing pressure. 

Definitely, the most spectacular situation occurs in the pressure range $p_c^\star < p < p_c$ (see Fig.~\ref{field} c). As discussed above, at zero field the opening of a superconducting gap on large parts of the Fermi surface, impedes the development of long range magnetic order inside the superconducting state. However, by applying a magnetic field parallel to $ab$ plane, when the superconducting state is weakend, antiferromagnetic order can reenter inside the superconducting state \cite{Park2006,Knebel2006}. Thus the opening of the superconducting gap does not avoid the formation of long range magnetic order if an external field is applied. In CeRhIn$_5$ the antiferromagnetic order is preserved above the upper critical field line $H_{c2} (T)$ and it seems that no anomaly in $H_{c2} (T)$ appears at the intersection of the $H_{c2} (T)$--$H_m (T)$ lines (see Fig.\ref{field} c). In this region of the phase diagram, magnetism and superconductivity seem to live peacefully. 

For $p<p_c$ the critical field of magnetism is always very large and the magnetic transition could be observed up to fields of $H=16$~T without any indication of a metamagnetic transition. Contrary, for pressure $p>p_c$ no indication of re-entrant antiferromagnetism is observed (see Fig.~\ref{field} d) ). The collapse of antiferromagnetism induced under magnetic field coincided with the critical pressure $p_c$.

\begin{figure}[tb]
\begin{center}
\includegraphics[width=.8\hsize,clip]{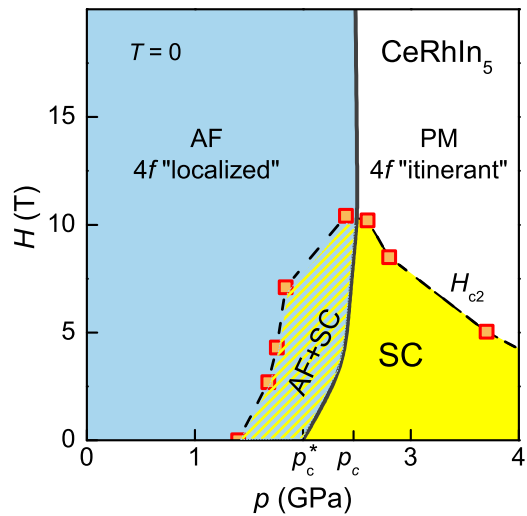}
\caption{\label{fermi_surface} ($H, T$) phase diagram of CeRhIn5 at $T=0$ indicating the Fermi surface topology in the different states of the phase diagram. The boundary between the localized Fermi surface (localized description of the 4$f$ electron) and of the itinerant paramagnetic phase (itinerant description of the 4$f$ electron) is indicated by thick black line. One yet unsolved question is the Fermi surface topology in the AF+SC state with the strong interplay between antiferromagnetism and superconductivity. One can speculate that at $H = 0$ an itinerant Fermi surface persists down to $p_c^\star$ as indicated.}
\end{center} 
\end{figure}

In CeRhIn$_5$ the $p_c$ corresponds also to an abrupt change of the de Haas van Alphen (dHvA) signal observed at high magnetic fields $H>8$~T \cite{Shishido2005}. From such measurements the topology of the Fermi surface and the effective mass of the charge carriers on the extremal orbits of the Fermi surface can be determined. Figure~\ref{fermi_surface} shows schematically the ($H, T$) phase diagram of CeRhIn$_5$ at $T=0$.
It has been shown that at ambient pressure the topology of the Fermi surface is almost identical to that of LaRhIn$_5$ which has no $f$ electron in the $4f$ shell, thus the $f$ electron in CeRhIn$_5$ looks localized and the large molecular field leads to recover the Fermi surface of LaRhIn$_5$. The dHvA signal changes abruptly as function of pressure at $p_c$. Above $p_c$ the frequencies of the orbits are shifted to higher fields indicating an increase of the Fermi volume which is at high pressure comparable to that of CeCoIn$_5$. Here the $f$ electron seem to be itinerant and takes part on the charge transport. The two-dimansional cylindrical character of the Fermi surface is maintained. Furthermore is has been shown that the cyclotron masses of the main dHvA branches increases steeply close to $p_c$ \cite{Shishido2005}.  
The driving force of this abrupt change of the Fermi surface is still under debate and a key ingredient maybe a possible pseudo-critical valence transition \cite{Miyake2006}, or some local quantum criticality at $p_c$ \cite{Harrison2004, Shishido2005, Park2008d}.

When superconductivity collapses above $H_{c2}$ it is clear that the Fermi surface change occurs at $p_c$ (see Fig. \ref{fermi_surface}). 
One can also speculate that at $H=0$ an itinerant Fermi surface persists down to $p_c^\star$ and under application of field the this change of the Fermi surface follows a line of first order transitions up to $p_c$ and that pure bulk superconductivity occurs only on the paramagnetic $4f$ itinerant Fermi surface. An yet unresolved question is the topology of the Fermi surface in the field induced AF+SC coexistence regime below $p_c^\star$. In this pressure and field regime no measurements exist. The hope in the future is to determine directly the Fermi surface in the AF+SC phase thanks to de Haas van Alphen oscillations just below $H_{c2}$.

What is the origin of the field induced AF+SC state? The first reason is that magnetic field weakens more strongly SC than AF up to $p_c$ and thus there will be a magnetic field $H^\star$  where in the pressure window ($p_c^\star, p_c$)
$T_c (H^\star)$ equals again $T_N (H^\star)$. Furthermore, the coexistence will occur at lower field since vortex cores are created. 
The idea is that antiferromagnetism may nucleate inside the vortex cores where superconductivity is expelled. Such a scenario has been discussed in more detail in the framework of the so-called SO(5) theory developed to explain the phase diagram of high $T_c$ superconductors (for a review see Ref.~\onlinecite{Demler2004}). In this theory the the antiferromagnetic order parameter and the superconducting order parameter are unified in a five dimensional vector called superspin which has magnetic and superconducting components. One fundamental prediction of this model is that a superconducting vortex can have an antiferromagnetic vortex core \cite{Arovas1997} and furthermore generate homogeneous AF over the whole sample.  Thanks to dynamical spin-density fluctuations extending even far from the vortex core a microscopic coexistence is created, in difference to a more static model where the magnetism is limited just to the vortex core \cite{Demler2001, Zhang2002}. It has been shown that under magnetic field the induced antiferromagnetic moment should increase with the applied field or the number of vortices.

\section*{Superconducting properties}

In the following we want to discuss superconducting properties of CeRhIn$_5$. As mentioned above, the remarkable feature is the abrupt suppression of the antiferromagnetic order when superconductivity sets in above $T_N$ at the pressure $p_c^\star$. Furthermore, it has to be noticed that $T_c (p)$ has a smooth pressure dependence through $p_c$.  What is the difference between the different superconducting phases? 

\begin{figure}[tb]
\begin{center}
\includegraphics[width=.8\hsize,clip]{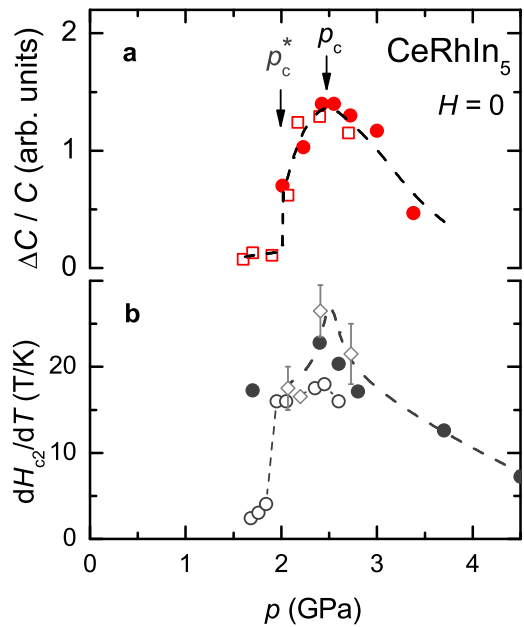}
\caption{\label{deltaC} \textbf{a}) Specific heat jump $\Delta C/C$ at the superconducting transition as function of pressure. (Different symbols corresponds to different experiments. (It should be noted that in the ac calorimetry it is not possible to get absolute values of the specific heat as it is not possible to determine from these measurements alone the correct for the background of the specific heat coming from the pressure cell and the pressure medium.)   \textbf{b}) Pressure dependence of the initial slope of the upper critical field at the superconducting transition. Full circles are from the resistivity experiments \cite{Knebel2008}, open diamonds are from ac calorimetry \cite{Knebel2006},  open circles are taken from Ref.~\onlinecite{Park2008b}, the pressure of Ref.~\onlinecite{Park2008b} has been normalized to our experiments.}
\end{center} 
\end{figure}

Figure \ref{deltaC} a) shows the pressure dependence of the specific heat anomaly $\Delta C/C$ at the superconducting transition at zero magnetic field. Two points are remarkable: firstly, the superconducting anomaly in the AF+SC state below $p_c^\star$ the specific heat anomaly at the superconducting transition is very small and increases step-like at $p_c^\star$. Secondly, at the critical pressure $p_c$ 
 $\Delta C/C$ shows a pronounced maximum and decreases to higher pressures. For a conventional BCS-like superconductor in the weak coupling limit, the jump at the superconducting transition has an universal value, independent of $T_c$. (From our ac calorimetry data it is not possible to get absolute values of the height of the jump, as it is not possible to estimate the addenda contribution to the specific heat coming from the pressure cell and the pressure medium.) The fact that the specific heat anomaly below $p_c^\star$ is very small but also not at all of BCS molecular field like has been first been interpreted as indication that superconductivity may not be bulk, but only a small volume fraction shows bulk superconductivity \cite{Knebel2004}.
As we have stressed above the specific heat anomaly is even not BCS like. The euphorique interpretation given previously is that it coincides with a concomittant superconducting and incommensurate -- commensurate antiferromagnetic transition. 

There is no conclusive determination of the superconducting order parameter of the coexistence regime. In the NQR experiments a residual density of states is observed, which led to speculations of gap-less superconductivity \cite{Kawasaki2003, Fuseya2003}. Some indications of $d$ wave symmetry of the superconducting order-parameter had been obtained by an angular dependent specific heat experiment which show a fourfold modulation of the specific heat as function of angle \cite{Park2008c}, but it is not obvious whether the very low temperature limit has been achieved in that experiment \cite{Vorontsov2007a,Vorontsov2007b}. 
In the framework of SO(5) theory the magnetic and superconducting order parameter are strongly coupled that for the superconducting order parameter (the superconducting gap $\Delta_{SC}$) and the magnetic order parameter (the staggered magnetization ($S_q$)) the constraint $\mid \!\! \Delta_{SC}\!\! \mid ^2 + \mid \!\! S_q \!\! \mid ^2 = 1$ has to be fulfilled
and indeed, the behavior of the respective critical temperatures under pressure implies such a coupling. The specific heat anomaly at $T_c$ increases under pressure in the same way as the internal magnetic field $H_{int}$ at the In site observed in NQR decreases. 
The decrease of the magnetic moment has been observed directly in neutron scattering experiments \cite{Raymond2008}.
In Ref.~\onlinecite{Park2009} the entropy at the magnetic transition $T_N$ and superconducting transition $T_c$ is analyzed as function of pressure and it has been shown that in the same way as the entropy at $T_N$ decreases the entropy at $T_c$ increases with pressure what shows the immediate coupling of both orders. 

Above $p_c^\star$ it is well accepted in the literature that CeRhIn$_5$ is a $d$-wave superconductor as the non magnetic compounds CeCoIn$_5$ and CeIrIn$_5$. This is mainly concluded from the temperature dependence of the nuclear spin relaxation rate $1/T_1$ which shows the famous $T^3$ dependence and the absence of the Hebel-Slichter peak. In this pressure region the jump of the specific heat is very large and of the same order than in CeCoIn$_5$ at zero pressure \cite{Park2008b}. This large value of the specific heat jump close to the critical pressure indicates the strong coupling
superconductivity directly linked to the strength of $T_c / T_A$.

Further insides on the superconducting properties can be obtained from the temperature dependence of the upper critical field. Generally, it is analyzed following the Werthammer-Helfland-Hohenberg model which takes into account the the orbital and the paramagnetic pair breaking effect \cite{Werthammer1966}. The orbital limiting field $H_{orb} (T) = \Phi_0 /2\pi \xi ^2$ is given by the fields at which vortex
cores starts to overlap ($\Phi_0$ is the flux quantum and $\xi$ the superconducting coherence length) and thus the superconducting state is suppressed. The  paramagnetic limiting field originates in the Zeeman splitting of the single electron energy levels. Superconductivity collapses if the applied field is stronger than the binding energy of the cooper pairs.

At low temperatures the upper critical field in CeRhIn$_5$ is completely dominated by the paramagnetic limitation.  
But close to $T_c$, the orbital limitation is always the dominant mechanism (the paramagnetic
limitation has infinite slope at $T_c$), so that the initial slope of $H_{c2}$ at $T_c$ is a good measure of the average Fermi
velocity in directions perpendicular to the applied field $H_{c2}' = ( d H_{c2})/d T)_{T = T_c} \approx T_c / v_F^2$. Thus from the initial slope it should be possible to estimate the effective mass of the charge carriers. The pressure dependence of the initial slope is shown in Fig. \ref{deltaC} b). Obviously, two very pronounced phenomena can be observed. In the AF+SC state the initial slope as detected in the ac calorimetry \cite{Park2008b} increases gradually with pressure up to $p_c^\star$, but at $p_c^\star$ a step increase of the initial slope appears. Thus in low magnetic fields an abrupt change in the electronic properties must occur at $p_c^\star$ at the border to the pure superconducting state. Looking at the properties of the Fermi surface this abrupt increase can be interpreted in that way that the average Fermi velocity decreases strongly or in terms of the effective mass of the charge carriers a jump of the average effective mass which is connected to the change of the Fermi surface from 4f localized to 4f itinerant for $H = 0$ at $p_c^\star$. But it should be noted that the heavy electrons feel the hidden quantum critical point at $p_c \approx 2.5$ GPa too. Here, the initial slope has a rather pronounced maximum which is even more accentuated than the maximum in $T_c (p)$. The maximum of the initial slope corresponds to a maximal in the effective mass. This reflects itself again in the superconducting properties. To analyze completely the upper critical field curve, at least above $p_c^\star$, it is necessary to take into account strong coupling effects (in standard BCS theory the electron phonon coupling, here the coupling between electrons and the fluctuations responsible for the pairing). For CeRhIn$_5$ we found a strong coupling parameter of $\lambda = 2.2$ close to the critical pressure, thus the mass enhancement due to the superconducting coupling is $ m_{sc}^\star/{m_b} = \lambda +1 \approx 3.2$. (In $m_{sc}^\star$ only those fluctuations are included which contributes to the superconducting pairing, $m_b$ is the band mass of the itinerant heavy quasiparticles.) Compared to the pressure variation of the effective mass as determined from the transport measurements at high pressure this is a rather smooth variation (e.g. the $A$ coefficient of the resistivity at high fields when superconductivity is suppressed varies by two orders of magnitude).  De facto the strong coupling constant changes under pressure but its changing is not directly linked to the pressure variation of $m^\star$ as $m_b$ itself can change drastically under pressure. 

\section*{Conclusions and generalities}

\begin{figure*}[tb]
\begin{center}
\begin{minipage}{0.49\hsize}
\begin{center}
\includegraphics[width=.9\hsize,clip]{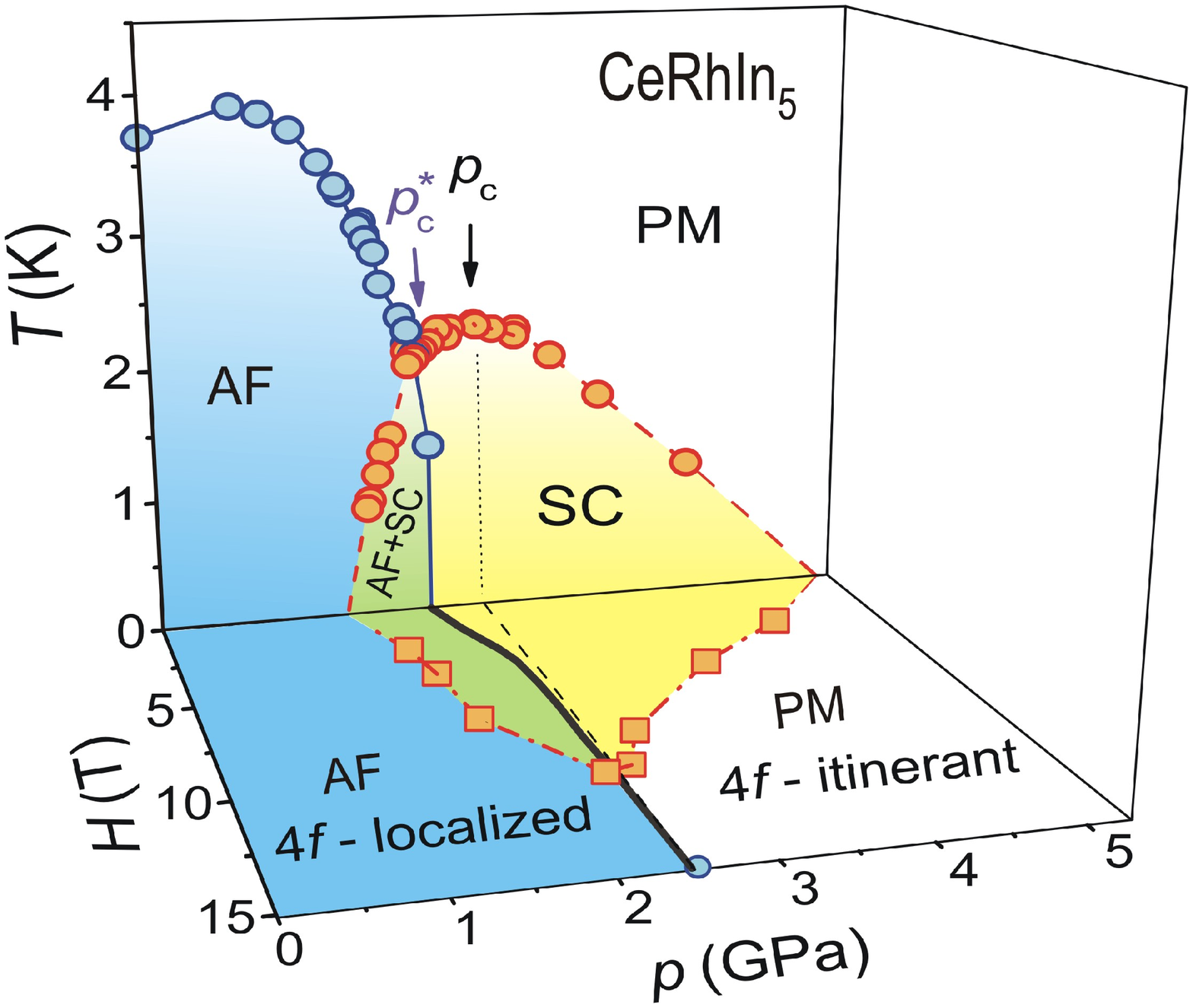}
\end{center} 
\end{minipage}
\begin{minipage}{0.49\hsize}
\begin{center}
\includegraphics[width=.8\hsize,clip]{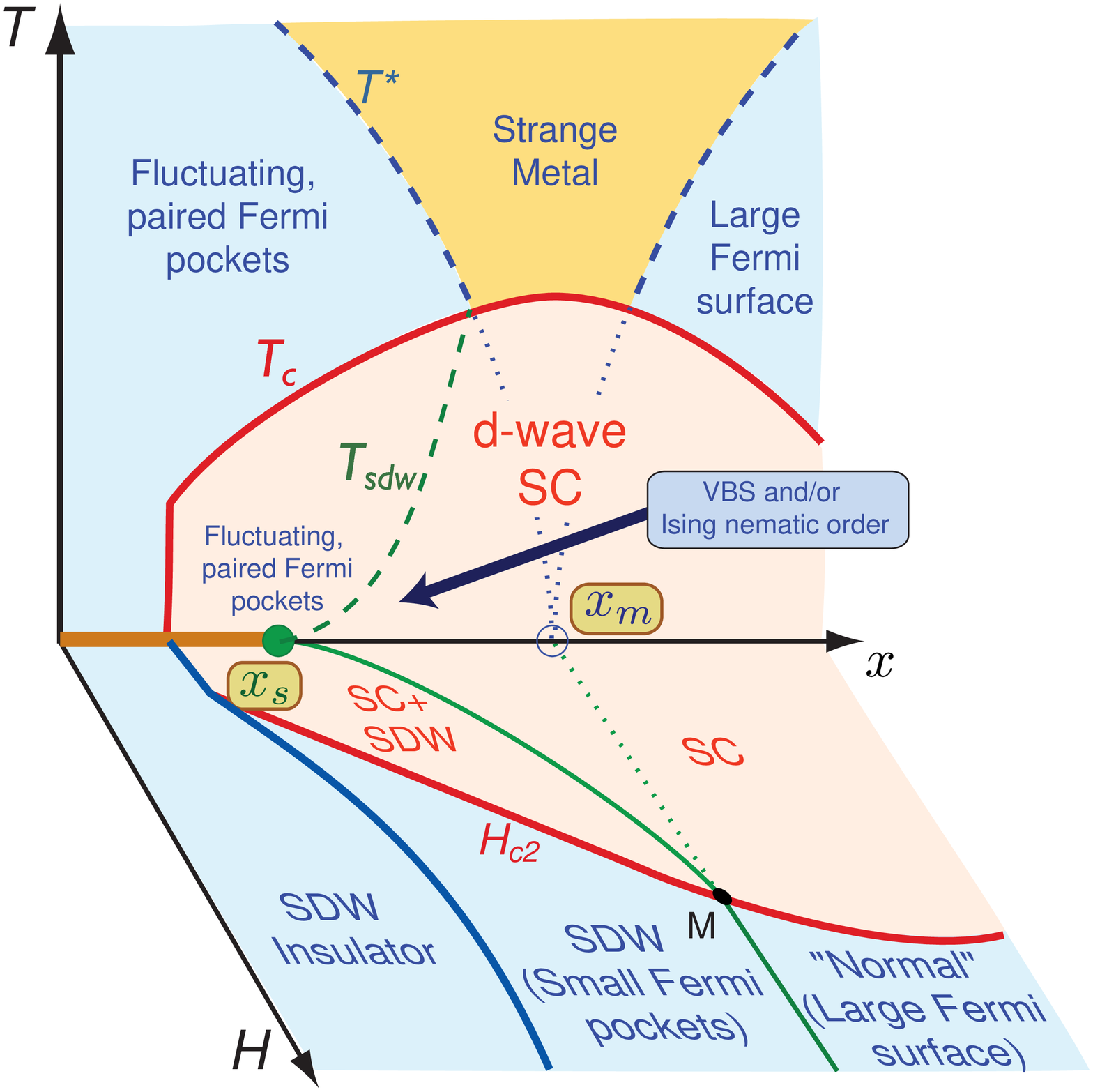}
\end{center} 
\end{minipage}
\caption{\label{Sachdev}(Left panel) Combined temperature, pressure and field $H\perp c$ phase diagram of CeRhIn$_5$  with antiferromagnetic (blue), superconducting (yellow), and coexistence AF+SC (green) phases. The thick black line in the $H-p$ plane indicates the proposed line where the Fermi surface changes from 4$f$ "localized" (small Fermi surface and topology comparable to LaRhIn$_5$), to 4$f$ "itinerant" (large Fermi surface as in CeCoIn$_5$).  (Right panel) Proposed phase diagram of the high $T_c$ cuprates showing the interplay between superconductivity (SC), spin density order (SDW), and Fermi surface configuration as function of carrier density ($x$), temperature ($T$), and magnetic field $(H)$ perpendicular to the CuO$_2$ layers (taken from S. Sachdev, Ref. \onlinecite{Sachdev2009}). }
\end{center} 
\end{figure*}

CeRhIn$_5$ is an exciting heavy-fermion system which allows to study in detail the interplay of long range antiferromagnetic order and superconductivity. Due to the fact that in this system the transition temperatures of antiferromagnetism and superconductivity are of the same order of magnitude a precise determination of the high pressure phase diagram could be realized. The combined $p, H, T$ phase diagram is plotted in Fig. \ref{Sachdev}. Under pressure and magnetic field three different regions appear, purely antiferromagnetic AF, a coexistence regime AF+SC, and a purely superconducting phase SC above $p_c^\star$. Thus as function of pressure at zero field the series of phases is AF -- AF+SC -- SC. 
In the pressure region between $p_c^\star$ and $p_c$ under application of magnetic field magnetic order is induced and one can observe the cascades of phases SC -- AF+SC -- AF with increasing field. There seems to be a difference between the pressure induced  AF+SC  and field induced AF+SC  states. When the AF state is established first under pressure at $H =0$, the measured specific heat jump and the initial slope of the upper critical field are still small and the superconducting transition at $T_c$ may coincide with a magnetic transition from an incommansurate to a commansurate structure. In contrast, when above $p_c^\star$ first superconductivity appears on cooling, the field induced AF state seems to have no influence on the SC properties such as $H_{c2} (T)$. Up to now no microscopic measurements exist in the pressure range from $p_c^\star$ to $p_c$ under magnetic field, obviously the field induced magnetic state is directly connected to the appearance of vortices in the mixed state of the superconductor.

Here we did not discuss the relation of the field induced AF+SC phase with the low temperature -- high magnetic field state in CeCoIn$_5$ \cite{Bianchi2003b,Matsuda2007}. This phase was first discussed being a Fulde-Ferrel-Larkin-Ovchinnikov (FFLO) state which is a novel superconducting state appearing under high magnetic fields in strongly Pauli limited superconductors characterized by the formation of Cooper pairs with non-zero total momentum \cite{Fulde1964, Larkin1964}. However, recently it could be shown by neutron scattering \cite{Kenzelmann2008}, and also by NMR \cite{Young2007, Koutroulakis2008}, that inside this state long range magnetic ordering coexist with superconductivity. In difference to the field induced AF+SC  state in CeRhIn$_5$ the novel phase in CeCoIn$_5$ appears only inside the superconducting state what indicates the strong coupling of both order parameters in that compound. Future measurements in CeRhIn$_5$ at very low temperatures just in the vicinity of the critical pressure $p_c$ have to address the possible appearance of such a novel superconducting state as well.

Finally we want to compare CeRhIn$_5$ to the case of high $T_c$ superconductors. 
The striking point is the similarity between the phase diagram of CeRhIn$_5$ and the recent proposal to clarify the situation in high $T_c$ cuprate superconductors \cite{Sachdev2009}. Both phase diagrams are shown in Fig.~\ref{Sachdev}, for CeRhIn$_5$ as function of temperature $T$, pressure $p$ and magnetic field $H$ applied in the basal plane, for the cuprates the variables are $T$, carrier concentration $x$ and magnetic field applied perpendicular to the CuO$_2$ layers. In zero magnetic field the onset of superconductivity hides the magnetic quantum critical point at $p_c$ and $x_m$ for the cuprate. But this underlying critical point determines the normal state properties leading to a "strange" metal domain.  In CeRhIn$_5$ it will be equivalent to the non Fermi liquid regime in the broad pressure window around $p_c$, for the cuprates it manifests itself in the linear temperatur dependence of the resistivity near optimal doping. The maximum of $T_c$ appears  at the critical pressure $p_c$ for CeRhIn$_5$ or for the concentration $x_m$ in case of the cuprates when the ordering temperature collapses to zero. (The temperarture of  the pseudo-gap $T^\star$ corresponds to antiferromagnetic transition temperature $T_N$ in CeRhIn$_5$.) In absence of superconductivity for $x\geq x_m$ the ground state of the cuprates would be  a paramagnetic Fermi liquid and for $x < x_m$ the ground state will be a spin density wave (SDW). As in CeRhIn$_5$ the onset of superconductivity with the appearance of a large superconducting gap impedes the formation of a SDW state just below $x_m$ and the SDW quantum critical point is shifted from $x_m$ to $x_s$ \cite{Moon2009}. For CeRhIn$_5$ the shift of the antiferromagnetic quantum critical point is from $p_c$ to $p_c^\star$ in zero magnetic field. 

At magnetic fields higher than the superconducting upper critical field $H_{c2}$ a Fermi surface instability is observed 
in both systems. As discussed above, in CeRhIn$_5$ the change at $p_c$ is from a small Fermi surface where the 4$f$ electrons look localized to a large Fermi surface where the 4$f$ electrons look itinerant \cite{Shishido2005}. For high $T_c$ cuprates the change of the Fermi surface at high fields is through $x_m$ with a large Fermi surface in the overdoped regime $x \geq x_m$ and small Fermi surface pockets in the underdoped regime $x < x_m$ \cite{Hussey2003,Vignolle2008,Doiron-Leyraud2007,Yelland2008,Bangura2008,Sebastian2008}. (At very low carrier concentration the ground state of the high $T_c$ superconductors is a antiferromagnetic Mott insulator.)

Under magnetic field, the weakness of $T_c$ as well as the creation of vortices favors the re-entrance of magnetic order in the superconducting state. For different high $T_c$ materials this has been experimentally shown by  neutron scattering experiments \cite{Lake2001, Lake2002, Kang2003, Chang2009}, NMR experiments \cite{Mitrovic2001, Kakuyanagi2003, Mitrovic2003}, or $\mu$SR \cite{Miller2002, Miller2006}. The boundary between SC+SDW and SC phase moves from $x_s$ to $x_m$ (M-point) when $H$ reaches $H_{c2} (x_m)$. For carrier concentrations $x_s < x < x_m$ a magnetic field scans will lead to the cascade of phases SC, SC+SDW, SDW. Thus the line between the points $(x_s, H=0), (x_m, H_M)$ in the $x, H$ plane at $T =0$ between SDW+SC and the pure superconducting phases delimits two different regimes linked in the SDW+SC phase to small Fermi surface pockets (left of the line) and in the sole SC phase at the right with a large Fermi surface. For CeRhIn$_5$ an analogue scenario seems valid in the pressure range from $p_c^\star$ to $p_c$. 

Despite difference in the basic interactions, CeRhIn$_5$ and the cuprate superconductors appear quite similar with the common points of hidden quantum criticality by superconductivity and the resurrection of magnetism under magnetic field. 

The discovery of the Ce-115 family with CeRhIn$_5$ showing a strong interplay of magnetism and superconductivity characterized by similar values of the maxima of their critical temperatures $T_N^{max} \approx T_c^{max}$ has given an unique opportunity to precise the pressure and magnetic field domain of coexistence of antiferromagnetism and superconductivity. As underlined, CeRhIn$_5$ gives now a sound basis for the understanding of unconventional superconductivity and its link to quantum criticality. By the comparison with the high $T_c$ superconductors we hope to have convinced the reader that 
impact of studies on competing phenomena has to be strong in the wide and rich domain of strongly correlated fermionic systems which goes from $^3$He liquid to heavy fermions, organic conductors and high $T_c$ cuprates. 

\subsection*{Acknowledgements}
We thank J.-P. Brison and S. Raymond for collaborations on CeRhIn$_5$. 
We acknowledge S. Sachdev for the phase diagram of high $T_c$ superconductors and comments. This work has been financially supported by the French ANR programmes IceNET, ECCE and DELICE.


%


\end{document}